**Krzysztof Moskwa**
Biblioteka Główna i OINT
Politechnika Wrocławska

**Piotr Rossa**
Biblioteka Uniwersytecka we Wrocławiu
Uniwersytet Wrocławski

# Rozwój bibliotek cyfrowych i repozytoriów elektronicznych na Dolnym Śląsku w latach 2004-2008

**STRESZCZENIE**

W referacie przedstawiono biblioteki cyfrowe i repozytoria elektroniczne funkcjonujące na Dolnym Śląsku w latach 2004-2008. Scharakteryzowano ogólnie ich zawartość i wielkość, zaprezentowano standardy i systemy zarządzania kolekcjami cyfrowymi oraz omówiono uwarunkowania prawne towarzyszące zarządzaniu zasobami cyfrowymi. Wskazano możliwości wykorzystania kolekcji cyfrowych w badaniach naukowych realizowanych w regionie.

**SUMMARY**
DEVELOPMENT OF DIGITAL LIBRARIES AND ELECTRONIC REPOSITORIES
IN LOWER SILESIA IN YEARS 2004-2008

In following elaboration were presented digital libraries and electronic repositories operating in Lower Silesia region in years 2004-2008. General description of character and size of their collections was presented, as well as standards and methods of digital collections management and juridical aspects of this management. Potential of usage of digital collections in regional scientific researches was described.

**Krzysztof Moskwa**
Biblioteka Główna i OINT
Politechnika Wrocławska

**Piotr Rossa**
Biblioteka Uniwersytecka we Wrocławiu
Uniwersytet Wrocławski

# Rozwój bibliotek cyfrowych i repozytoriów elektronicznych na Dolnym Śląsku w latach 2004-2008

## STRESZCZENIE

W referacie przedstawiono biblioteki cyfrowe i repozytoria elektroniczne funkcjonujące na Dolnym Śląsku w latach 2004-2008. Scharakteryzowano ogólnie ich zawartość i wielkość, zaprezentowano standardy i systemy zarządzania kolekcjami cyfrowymi oraz omówiono uwarunkowania prawne towarzyszące zarządzaniu zasobami cyfrowymi. Wskazano możliwości wykorzystania kolekcji cyfrowych w badaniach naukowych realizowanych w regionie.

## CEL I PRZEDMIOT

Celem referatu jest ustalenie liczby projektów związanych z tworzeniem bibliotek cyfrowych (bc) i repozytoriów elektronicznych (re) na Dolnym Śląsku. Ponadto określenie ich specyfiki, wielkości, tempa rozwoju i ogólnego poziomu zainteresowania ze strony czytelników. Zwrócono również uwagę na złożone uwarunkowania prawne towarzyszące zarządzaniu projektami udostępniającymi zasoby cyfrowe. Istotnym celem jest także promowanie projektów sprzyjających powszechnej komunikacji naukowej i kulturalnej, „otwartego" oprogramowania i standardów oraz promowanie bibliotek i ich działań na rzecz środowiska naukowego i mieszkańców regionu. Daty graniczne wyznacza utworzenie pierwszej biblioteki cyfrowej, tj. *Biblioteki Cyfrowej Politechniki Wrocławskiej* w 2004 roku oraz schyłek roku kalendarzowego poprzedzającego wygłoszenie referatu. Dane liczbowe prezentowane w referacie zostały zaczerpnięte ze statystyk lub informacji zawartych na stronach przywołanych projektów [stan na 31.12.2008], a w przypadku *Repozytorium Eny Politechnika Wrocławska* także z wywiadu przeprowadzonego z jego administratorem.

## WPROWADZENIE

Rozwój technologii informacyjno-komunikacyjnych warunkuje również transformacje w otoczeniu nauki wpływając na zmiany w sposobach publikowania literatury naukowej przez wydawców, ale również ich gromadzenia, opracowania, przetwarzania i udostępniania przez biblioteki działające na rzecz środowiska naukowego. Działalność bibliotek widziana oczyma czytelników przenosi się w coraz większym stopniu z fizycznych budynków do przestrzeni wirtualnej, gdzie odnaleźć można już nie tylko informacje o zbiorach gromadzonych w bibliotekach (katalogi online), ale również same zbiory: e-czasopisma, e-książki, bazy danych, a w ostatnich latach licznie przyrastające kolekcje obiektów elektronicznych organizowanych i zarządzanych przez biblioteki. Biblioteki oferujące zarówno zbiory tradycyjne, jak i elektroniczne, określane są mianem bibliotek hybrydowych, które w sposób racjonalny kształtują przestrzeń informacyjną bibliotek.

Jedną z najbardziej ogólnych definicji opisujących przedmiot niniejszego referatu, czyli kolekcje obiektów elektronicznych dostępnych w Internecie, przynosi Wikipedia.

Repozytorium (łac. *repositorium*) oznacza „*miejsce uporządkowanego przechowywania dokumentów, z których wszystkie przeznaczone są do udostępniania... Niegdyś szafa na księgi i akta* urzędowe. *Dziś termin stosowany również w odniesieniu do najrozmaitszych zasobów cyfrowych (baz danych, zbioru pakietów czy kodów źródłowych), np. w Internecie*" [23].

Repozytoria elektroniczne podzielić można na dwie podstawowe grupy:
- repozytoria dziedzinowe, rozpowszechnione wśród naukowców reprezentujących dziedziny dynamicznie się rozwijające, skłonnych do szybkiego dzielenia się wynikami badań, np. *arXiv* [http://arxiv.org/] z zakresu fizyki, matematyki, informatyki
- repozytoria instytucjonalne „*służą archiwizacji intelektualnego dorobku instytucji*" [9] i wyróżnić można:
  - repozytoria instytucjonalne jednej instytucji
  - repozytoria ponadinstytucjonalne łączące grupę instytucji tego samego typu lub wyodrębnione ze względu na lokalizację geograficzną, np. regionalne bc.

Repozytoria mogą stanowić też część bibliotek cyfrowych [9] [13], które nazywane są również bibliotekami internetowymi, sieciowymi, elektronicznymi, wirtualnymi [8] [13] [14] [22].

Starsze definicje bibliotek cyfrowych [2] kładą nacisk na aspekt technologiczny (informacja elektroniczna, sposób dostępu), aspekt merytoryczny (wyselekcjonowana i uporządkowana informacja) i aspekt ekonomiczny (racjonalny dobór dokumentów).

Na problemy definicyjne w języku polskim zwracano uwagę już w latach 90-tych XX wieku [3] [14], współcześnie tamte rozważania nie straciły na aktualności, a bibliotekę cyfrową rozumianą jako „*cyfrowe zasoby, cyfrowe opracowanie i cyfrowe udostępnianie*" przywołują bardziej aktualne publikacje [17]. Jednak podnoszone są też argumenty, czy mamy do czynienia z wieloma rozproszonymi bc, czy dzięki technologiom i standardom ułatwiającym komunikację, jedną uniwersalną bc [8]. W takim ujęciu odróżnienie zbiorów bc od ogółu informacji dostępnych w Internecie sprowadza się do celowego i profesjonalnego przygotowania, opracowania oraz zabezpieczenia danych.

Powyższe przykłady pokazują, że znaczenie terminu biblioteka cyfrowa będzie ewoluować wraz ze zmieniającym się kontekstem organizacyjnym, technologicznym i prawnym.

Bc i re mogą zawierać obiekty cyfrowe wytworzone w tej postaci lub poddane procesowi digitalizacji, przetworzone z postaci analogowej (np. książki, czasopisma, obrazy, zdjęcia, dane dźwiękowe i dźwiękowo-wizualne).

Podstawowe formaty dokumentów występujące w bc i re, to m.in. dla dokumentów tekstowych: HTML, SGML, XML, PDF, TXT, MS Office, Open Office, ODF, dla grafiki: JPEG, GIF, TIFF, PNG, PhotoCD, DjVu, SVG; dla muzyki: MIDI, WAVE, MP3, Real Audio, WMA; dla multimediów: Real Media (RM), MPEG, Flash (SWF), SMIL [2] [20].

Praktyka tworzenia krajowych zasobów cyfrowych wykształciła dwa wyraźnie zróżnicowane modele organizowania i zarządzania kolekcjami obiektów elektronicznych oraz ich odmienne role [13]. W tworzenie bc zaangażowane są przede wszystkim instytucjonalne biblioteki, re powstają również poza bibliotekami. Bc działające w środowisku akademickim oprócz opracowań o charakterze naukowym prezentują materiały o znaczeniu archiwalnym, historycznym czy literackim, co podkreśla ich funkcje dokumentujące i kulturotwórcze. Repozytoria zorientowane są niemal wyłącznie na prace naukowe.

## USTALENIE LICZBY PROJEKTÓW

W wyniku analizy przeprowadzonej [stan na 31.12.2008] w dwóch globalnych serwisach agregujących informacje o bibliotekach cyfrowych i repozytoriach (*OpenDOAR*, *ROAR*), a także w uniwersalnym katalogu rejestrującym biblioteki cyfrowe i repozytoria spełniające standardy otwartych informacji o archiwach (*OAIster*) oraz w krajowym projekcie integrującym informacje o zasobach bibliotek cyfrowych (*FBC*), wyłoniono listę krajowych projektów, która stała się podstawą ustalenia bibliotek cyfrowych i repozytoriów tworzonych i zarządzanych przez instytucje naukowe i kulturalne Dolnego Śląska.

Poniższe katalogi pełnią również funkcję wyszukiwarek zasobów cyfrowych umożliwiających prowadzenie rozproszonych poszukiwań w zarejestrowanych w nich projektach, w oparciu o informację zwartą w metadanych gromadzonych obiektów cyfrowych.

*OpenDOAR* [http://www.opendoar.org/] – *Directory of Open Access Repositories* rejestruje 1 298 projektów z całego świata, Polskę reprezentuje 13 projektów (1%), z tego na Dolnym Śląsku zlokalizowano 3, co stanowi 23% krajowych przedsięwzięć.

*ROAR* [http://roar.eprints.org/] – *Registry of Open Access Repositories* odnotowuje 1 239 projektów z całego świata, Polskę reprezentuje 7 projektów (0.6%), z tego Dolny Śląsk reprezentuje 1 projekt, stanowi to 14% krajowych przedsięwzięć.

*OAIster* [http://www.oaister.org/] rejestruje projekty spełniające standardy OAI-PMH (*Open Archives Initiative Protocol for Metadata Harvesting*), w liczbie 1 054, udostępniających, zarówno licencjonowane treści płatne, jak i ogólnodostępne zasoby. Polskie biblioteki cyfrowe i repozytoria odnotowane w *OAIster* mają charakter niekomercyjny i umożliwiają powszechny dostęp do zasobów. Spośród 19 krajowych projektów (1,8%) na Dolnym Śląsku funkcjonują 3 (16%).

*FBC* [http://fbc.pionier.net.pl/] – *Federacja Bibliotek Cyfrowych* umożliwia jednoczesne przeszukiwanie zasobów 30 krajowych bibliotek cyfrowych spełniających standardy OAI-PMH. Dolny Śląsk jest reprezentowany przez 3 projekty (10%).

Po wykluczeniu powtarzających się wyników ustalono, że na Dolnym Śląsku realizowane są cztery projekty związane z tworzeniem zasobów cyfrowych:
- *Biblioteka Cyfrowa Uniwersytetu Wrocławskiego* (*BCUWr*),
- *Dolnośląska Biblioteka Cyfrowa* (*DBC*),
- *Jeleniogórska Biblioteka Cyfrowa* (*JBC*),
- *Repozytorium Eny Politechnika Wrocławska* (*Eny PWr*).

## CHARAKTERYSTYKA PROJEKTÓW

Projekty realizowane na Dolnym Śląsku wykazują zróżnicowanie [Tab. 1] pod względem organizacyjnym i technologicznym. *BCUWr* i *Eny PWr* zarządzane są przez pojedyncze instytucje, *DBC* i *JBC* na podstawie porozumienia kilku lub kilkunastu współdziałających instytucji.

W *BCUWr*, *DBC* i *JBC* działa system płatny *dLibra* w repozytorium *Eny PWr* bezpłatna aplikacja *CDS Invenio*.

Tab. 1. Projekty realizowane na Dolnym Śląsku w latach 2004-2008

| Opis / Projekt | Rok utworzenia | Liczba obiektów [31.12.2008] | Liczba kolekcji | System | Status systemu | Format danych |
|---|---|---|---|---|---|---|
| *DBC / BC PWr* | 2004 | 1311 | 16 | *dLibra* | płatny | Dublin Core |
| *BCUWr* | 2005 | 18013 | 6 | *dLibra* | płatny | Dublin Core |
| *JBC* | 2007 | 703 | 9 | *dLibra* | płatny | Dublin Core |
| *Eny PWr* | 2008 | 276 | 4 | *CDS Invenio* | bezpłatny | MARC21 |

*Biblioteka Cyfrowa Uniwersytetu Wrocławskiego* [http://www.bibliotekacyfrowa.pl/]

Uroczyste otwarcie *BCUWr* nastąpiło 5 grudnia 2006 roku [19]. W czasie rocznego testowania platformy *dLibra*, poprzedzającego moment oficjalnego uruchomienia wprowadzono do *BCUWr* ponad 7 tysięcy pozycji, z których skorzystało prawie 960 tysięcy użytkowników [11].

*Biblioteka Cyfrowa UWr* została stworzona przez Bibliotekę Uniwersytecką we Wrocławiu w celu zachowania i popularyzacji bogatych zasobów dziedzictwa kulturowego zawartego w jej zbiorach. Umożliwia ona dostęp do cyfrowych kopii najcenniejszych zabytków kultury piśmienniczej, kolekcji dzieł sztuki oraz materiałów regionalnych. Zbiory *BCUWr* zostały podzielone na sześć kolekcji:
- Dziedzictwo kulturowe,
- e-Książki,
- Katalogi BUWr,
- Kolekcje specjalistyczne,
- Materiały edukacyjne,
- Regionalia.

Każda z kolekcji podzielona została na podkolekcje. W ramach kolekcji i podkolekcji „Dziedzictwo kulturowe" udostępnia się elektroniczne kopie: czasopism z XIX i z pierwszej połowy XX wieku, dokumentów życia społecznego, książek, rękopisów, starych druków. W zasobach cyfrowych reprezentowane są także zbiory bibliologiczne, graficzne, kartograficzne, oraz muzyczne.

„e-Książki" udostępniane są w całości lub we fragmentach w zależności od warunków licencji na jakich zostały przekazane przez wydawców. W kolekcji tej prezentowane są także e-booki wydawane w ramach *e-Wydawnictwa BUWr.*

„Kolekcje specjalistyczne" prezentują kopie cyfrowe wszystkich rodzajów dokumentów wybranych ze zbiorów Archiwum UWr, Muzeum UWr, Biblioteki Ogrodu Botanicznego UWr oraz Biblioteki Wydziału Prawa, Administracji i Ekonomii UWr (*Prawnicza i Ekonomiczna BC*). Kryterium doboru dzieł do tej kolekcji stanowi ich wartość merytoryczna lub historyczna.

Do „Materiałów edukacyjnych" zaliczono doktoraty, podręczniki, skrypty oraz inne materiały pomocne w procesie nauczania na Uniwersytecie Wrocławskim. Zastosowano tu ogólny podział nauk na ścisłe, przyrodnicze, społeczne i humanistyczne.

„Regionalia" obejmują cyfrowe kopie typów dokumentów wymienionych w kolekcji „Dziedzictwo kulturowe" dotyczących obszaru Śląska i Łużyc, ze szczególnym uwzględnieniem Dolnego Śląska i miasta Wrocławia. Wyróżniono tu trzy podkolekcje: Łużyce, Śląsk oraz Wrocław.

Użytkownik może przeprowadzić wyszukiwanie ogóle i zaawansowane oraz wyszukiwanie w indeksach (indeks tytułów, indeks twórców oraz indeks słów kluczowych).

Interfejs posiada, oprócz polskiej, cztery wersje językowe: angielską, czeską, francuską i niemiecką.

*BCUWr* działa na platformie *dLibra* 4.0, współdziała z protokołem OAI-PMH, dane opracowywane są w formacie Dublin Core na podstawie *ePoradnika redaktora zasobów cyfrowych* przygotowanego przez zespół pracowników Biblioteki Uniwersyteckiej we Wrocławiu. Aktualną wersją poradnika jest wersja z dnia 3 lipca 2008 roku, opublikowana na licencji Creative Commons (Uznanie autorstwa 2.5 Polska, CC-BY) [4].

Tab. 2. Zawartość i wykorzystanie *BCUWr* [stan na 31.12.2008]

| Rok<br>Aktywność | 2005 | 2006 | 2007 | 2008 |
|---|---|---|---|---|
| Łączna liczba obiektów | 18 | 7 103 | 1 0226 | 18 023 |
| Przyrost roczny obiektów | 18 | 7 085 | 3 123 | 7 797 |
| Liczba pobrań | 436 | 52 724 | 100 469 | 179 324 |
| Liczba wyszukiwań | 711 | 96 638 | 197 060 | 189 403 |
| Liczba odwiedzających | 6 385 | 924 323 | 697 267 | 876 191 |
| Średnia liczba pobrań na 1 obiekt | 24 | 7 | 10 | 10 |
| Liczba pobrań a liczba odwiedzin [%] | 7% | 6% | 14% | 20% |

Łącznie w analizowanym okresie [Tab. 2] w *BCUWr* zgromadzono 18 023 obiekty, stronę odwiedziło 2 504 166 czytelników, pobierając łącznie 332 953 dokumenty. Oznacza to, że jeden obiekt został średnio wykorzystany 18 razy, a niemal co ósme wejście na stronę *BCUWr* kończyło się pobraniem dokumentu przez czytelnika (13%).

*Dolnośląska Biblioteka Cyfrowa* [http://www.dbc.wroc.pl/]

Inicjatywą poprzedzająca powołanie *DBC* było utworzenie w 2004 roku *Biblioteki Cyfrowej Politechniki Wrocławskiej* (*BC PWr*) działającej na platformie *dLibra*, w strukturze organizacyjnej Biblioteki Głównej i OINT Politechniki Wrocławskiej. W 2005 roku współdziałanie Biblioteki Głównej i OINT z Zakładem Narodowym im. Ossolińskich oraz od strony technicznej z Wrocławskim Centrum Sieciowo-Superkomputerowych doprowadziło do przekształcenia *BC PWr* w *DBC* [15]. W 2006 roku założyciele oraz uczestnicy *DBC* powołali Konsorcjum Dolnośląskiej Biblioteki Cyfrowej. Zmianie uległ zatem status bc, z platformy instytucjonalnej stała się bc o charakterze regionalnym. Powstanie konsorcjum nie powołało nowego podmiotu prawa, a jedynie nową formę organizacyjną współpracy uczestników.

Z końcem 2008 roku Konsorcjum Dolnośląskiej Biblioteki Cyfrowej liczyło 15 uczestników instytucjonalnych, zarządzających własnymi kolekcjami:
- Politechnika Wrocławska,
- Zakład Narodowy im. Ossolińskich,
- Akademia Wychowania Fizycznego,
- Uniwersytet Ekonomiczny,
- Uniwersytet Przyrodniczy,
- Akademia Muzyczna,
- Kolegium Karkonoskie,
- Papieski Wydział Teologiczny,
- Wyższa Szkoła Oficerska Wojsk Lądowych,
- Akademia Medyczna,
- Akademia Sztuk Pięknych,
- Państwowa Wyższa Szkoła Teatralna,
- Dolnośląska Biblioteka Pedagogiczna,
- Politechnika Opolska,
- Wojewódzki Urząd Ochrony Zabytków we Wrocławiu,
- Środowisko Akademickie.

Instytucje współpracujące z *DBC* na innych zasadach niż członkowie konsorcjum mogą udostępniać swoje materiały w kolekcji „Środowisko akademickie".

Biblioteka udostępnia wersje elektroniczne zbiorów tradycyjnych posiadanych przez uczestników konsorcjum, do których wygasły autorskie prawa majątkowe. W przypadku utworów objętych pełną ochroną prawną, podstawą umieszczenia w *DBC* jest niewyłączna umowa licencyjna podpisana przez posiadacza praw majątkowych [15].

Selekcja materiałów do skanowania i publikowania opiera się na następujących kryteriach:
- aktualność – zbiory współczesne,
- unikatowość – stare druki, zbiory cenne, prestiżowe, bogato ilustrowane,
- tematyka – zbiory związane z Dolnym Śląskiem i Wrocławiem, zbiory z Narodowego Zasobu Bibliotecznego, zbiory naukowe zgodne z profilem działalności uczestników Konsorcjum,
- potrzeby czytelników – na podstawie wykorzystania zbiorów tradycyjnych,
- możliwości techniczno-sprzętowe – ograniczenia skanerów, format wersji papierowej, stan zachowania skanowanych materiałów [21].

*DBC* działa obecnie na platformie *dLibra* w wersji 4.0, współpracuje z protokołem OAI-PMH, dane opracowywane są w formacie opisu metadanych Dublin Core.

Od 2008 roku opracowanie opisu obiektów odbywa się na podstawie *Instrukcji tworzenia opisu metadanowego w formacie Dublin Core Metadata Element Set w Konsorcjum Dolnośląskiej Biblioteki Cyfrowej* [21].

*DBC* umożliwia wyszukiwanie proste i zaawansowane w oparciu o kilkanaście kryteriów oraz przeglądanie indeksów: tytułów, twórców i słów kluczowych, a także przeglądanie zawartości kolekcji. Intrefejs przygotowano w 7 wersjach językowych: polskiej, angielskiej, czeskiej, francuskiej, niemieckiej, rosyjskiej i ukraińskiej.

Tab. 3. Zawartość i wykorzystanie *DBC* [stan na 31.12.2008]

| Rok<br>Aktywność | 2004 | 2005 | 2006 | 2007 | 2008 |
|---|---|---|---|---|---|
| Łączna liczba obiektów | 50 | 144 | 358 | 811 | 1 311 |
| Przyrost roczny obiektów | 50 | 94 | 214 | 453 | 500 |
| Liczba pobrań | 1 182 | 34 258 | 63 185 | 343 120 | 520 850 |
| Liczba wyszukiwań | 4 450 | 78 240 | 139 427 | 253 194 | 240 713 |
| Liczba odwiedzających | 18 110 | 249 912 | 717 627 | 1 033 747 | 1 162 827 |
| Średnia liczba pobrań na 1 obiekt | 24 | 238 | 176 | 423 | 397 |
| Liczba pobrań a liczba odwiedzin [%] | 7% | 14% | 9% | 33% | 45% |

Łącznie w analizowanym okresie [Tab. 3] w *DBC* zgromadzono 1 311 obiektów, stronę odwiedziło 3 182 223 czytelników, pobierając łącznie 962 595 dokumentów. Oznacza to, że jeden obiekt został średnio wykorzystany 734 razy, a prawie co trzecie wejście (30%) na stronę *DBC* kończyło się pobraniem dokumentu przez czytelnika.

*Jeleniogórska Biblioteka Cyfrowa* [http://jbc.jelenia-gora.pl/]

*JBC* jest regionalną biblioteką cyfrową. Powstała ona z inicjatywy Książnicy Karkonoskiej w Jeleniej Górze, przy wsparciu finansowym Samorządu Województwa Dolnośląskiego. Oprócz Książnicy Karkonoskiej w przedsięwzięciu uczestniczą także:

- Archiwum Państwowe we Wrocławiu. Oddział w Jeleniej Górze,
- Dolnośląska Biblioteka Publiczna im. Tadeusza Mikulskiego we Wrocławiu,
- Dolnośląskie Centrum Doskonalenia Nauczycieli i Informacji Pedagogicznej. Biblioteka Pedagogiczna w Legnicy. Filia w Jeleniej Górze,
- Jeleniogórskie Centrum Kultury,
- Muzeum Karkonoskie.

Serwis został uruchomiony we wrześniu 2007 roku na platformie *dLibra* [20]. Biblioteka ta stawia sobie za cel prezentację i promocję rzadkich i cennych publikacji regionalnych: książek, czasopism, dokumentów życia społecznego, pamiętników Jeleniogórzan i biogramów ludzi wpisanych w historię miasta i regionu, map, grafiki oraz pocztówek związanych z Jelenią Górą i Dolnym Śląskiem, będących w posiadaniu Grodzkiej Biblioteki Publicznej i uczestników projektu.

Zbiory *JBC* zostały podzielone na następujące główne kolekcje:

„Archiwalia", „Czasopisma", „Dokumenty Życia Społecznego", „Film", „Ikonografia", „Książki", „Pocztówki", „Słownik Biograficzny Ziemi Jeleniogórskiej", „Varia".

Kolekcja „Archiwalia" to zbiór dokumentów historycznych związanych z regionem. „Czasopisma" to kolekcja oferująca dostęp do cyfrowych kopii regionalnych czasopisma archiwalnych. W kolekcji „Dokumentów Życia Społecznego" wyróżniono podkolekcje „Plakaty i afisze" oraz „Wspomnienia". „Film" to kolekcja filmów dokumentalnych, rejestrująca wydarzenia związane z Ziemią Jeleniogórską i Jelenią Górą. „Ikonografia" prezentuje twórczość artystów plastyków związanych z regionem, wyróżniono podkolekcję: „Fotografia i Grafika".

W *JBC* czytelnik może znaleźć elektroniczne wersje cennych wydawnictw przechowywanych w Grodzkiej Bibliotece Publicznej w Jeleniej Górze oraz bibliografię zawartości Rocznika Jeleniogórskiego (kolekcja „Książki"). Prócz tego do dyspozycji ma także podkolekcję bibliografii regionalnych.

Ważną kolekcję stanowią kopie cyfrowe pocztówek z regionu jeleniogórskiego (kolekcja „Pocztówki"). Ich oryginały przechowywane są w zbiorach Książnicy Karkonoskiej. W kolekcji tej wprowadzono podział na podkolekcje według kryterium geograficznego.

Kolekcja „Słownik Biograficzny Ziemi Jeleniogórskiej" zawiera biogramy osób nieżyjących związanych z Jelenią Górą i regionem.

„Varia" to kolekcja, która skupia w sobie materiały niemieszczące się w innych kolekcjach, ale warte opublikowania w wersji elektronicznej. Są to między innymi cenne informacje z prasy, wycinki, artykuły, opracowania dotyczące regionalnych atrakcji turystycznych, zabytków, obiektów, wydarzeń. Przeszukiwanie jest możliwe w oparciu o formularz ogólny i zaawansowany oraz przez indeksy (indeks tytułów, indeks twórców, indeks słów kluczowych). Czytelnik *JBC* może wybrać jedną z siedmiu wersji językowych interfejsu (polską, niemiecką, czeską, ukraińską, angielską, francuską i rosyjską).

*JBC* działa na platformie *dLibra* 4.0, która została zakupiona ze środków Ministerstwa Kultury i Dziedzictwa Narodowego. *JBC* współpracuje z protokołem OAI-PMH, dane opracowywane są w formacie Dublin Core.

Tab. 4. Zawartość i wykorzystanie *JBC* [stan na 31.12.2008]

| Rok<br>Aktywność | 2007 | 2008 |
|---|---|---|
| Łączna liczba obiektów | 434 | 703 |
| Przyrost roczny obiektów | 434 | 269 |
| Liczba pobrań | 14 229 | 44 760 |
| Liczba wyszukiwań | 5 601 | 20 141 |
| Liczba odwiedzających | 60 695 | 141 499 |
| Średnia liczba pobrań na 1 obiekt | 33 | 64 |
| Liczba pobrań a liczba odwiedzin [%] | 23% | 32% |

Łącznie w analizowanym okresie [Tab. 4] w *JBC* zgromadzono 703 obiekty, stronę odwiedziło 202 194 czytelników, pobierając łącznie 58 989 dokumentów. Oznacza to, że jeden obiekt został średnio wykorzystany 84 razy, a prawie co trzecie wejście (29%) na stronę *JBC* kończyło się pobraniem dokumentu przez czytelnika.

*Repozytorium Eny Politechnika Wrocławska* [http://zet10.ipee.pwr.wroc.pl/]

Repozytorium działa przy Wydziale Elektrycznym Politechniki Wrocławskiej, koncepcja jego utworzenia powstała w latach 2002-2005 [6], a faktyczne jego udostępnienie w Internecie miało miejsce w marcu 2008 roku. Repozytorium umożliwia autorom autoarchiwizację, różnego rodzaju prac publikowanych i niepublikowanych. Obecnie znajdują się w nim m.in. artykuły z czasopism, zarówno wersji preprint (przed recenzją naukową), jak i postprint (po opublikowaniu w czasopiśmie), propozycje książkowe, referaty konferencyjne, raporty, prace dyplomowe, materiały dydaktyczne oraz zdjęcia.

W repozytorium mogą umieszczać swoje utwory pracownicy, doktoranci i studenci Wydziału Elektrycznego, ale również inni autorzy z Uczelni i spoza niej, których tematyka prac jest zbieżna z profilem kolekcji. Warunkiem jest indywidualna rejestracji autora w repozytorium i udzielenie praw niewyłącznych na udostępnianie przezeń utworu, dopuszczalne są różne warianty licencji niewyłącznych [6]. Według informacji zawartej na stronie projektu, utwory zawarte w *Eny PWr* udostępniane są na licencji Creative Commons (Uznanie autorstwa - Użycie niekomercyjne - Na tych samych warunkach 2.5 Polska; CC-BY-NC-SA).

W początkowym okresie działalności repozytorium odnotowywało ok. 500-700 wejść miesięcznie, w sierpniu 2009 roku, liczba odwiedzin przekroczyła 1 000 [na podst. danych administratora].

Do końca 2008 roku zdeponowano w repozytorium 276 dokumentów podzielonych na cztery główne kolekcje („Artykuły i Preprinty", „Książki i Raporty"; „Dydaktyka", „Multimedia i Fotografie"), ok. 90% z nich stanowiły utwory o charakterze naukowym.

Korzyści, na jakie wskazują organizatorzy przedsięwzięcia, to głównie realizacja potrzeb własnego i zewnętrznego środowiska akademickiego i badawczego, zacieśnianie współpracy i tworzenie nowych grup badawczych. Ale również znaczący wzrost liczby cytowań, łatwiejsza dostępność publikacji, w tym dostęp do dorobku instytucji, zwiększona widzialność i rozpoznawalność instytucji oraz jej reklama. Ponadto budowanie prestiżu i wzrost pozycji instytucji w rankingach krajowych i międzynarodowych, przewodnictwo w dziedzinie rozpowszechniania wiedzy, i co ważne z ekonomicznego punktu widzenia, zmniejszenie kosztów klasycznego marketingu instytucji przy trwałych efektach. Przy okazji realizacji powyższych celów także ułatwione zarządzanie materiałami dydaktycznymi i szkoleniowymi oraz promocja „otwartych" inicjatyw [6].

## PLATFORMY I STANDARDY

Zaawansowane technologicznie projekty udostępniają zasoby cyfrowe poprzez dedykowane tego typu przedsięwzięciom platformy systemowe. Wykaz repozytoriów *OpenDOAR* [http://www.opendoar.org/find.php] odnotowuje ok. 70 zdefiniowanych rozwiązań wykorzystywanych przez biblioteki cyfrowe i repozytoria na całym świecie. Począwszy od statycznych zestawień w HTML do specjalistycznych aplikacji komercyjnych lub rozwijanych na otwartych licencjach.

Cztery projekty realizowane na Dolnym Śląsku wykorzystują dwie z platform stworzonych do tego celu. *BCUWr*, *DBC* i *JBC* działają w oparciu o system *dLibra*, a repozytorium *Eny PWr* wykorzystuje *CDS Invenio*.

*dLibra* [http://dlibra.psnc.pl/]

System utworzony i rozwijany przez Poznańskie Centrum Superkomputerowo-Sieciowe (PCSS) od 1999 roku. Koszt licencji obecnej wersji systemu (4.0) to 1 200 zł netto, pozwala ona na bezterminowe korzystanie z udostępnionej wersji bez ograniczeń funkcjonalnych. Koszt może wzrosnąć, jeśli zamawiający skorzysta z dodatkowych usług oferowanych przez PCSS, m.in. przeprowadzenia instalacji i konfiguracji, szkoleń, pomocy technicznej, migracji oprogramowania do wyższych wersji.

Domyślnym schematem opisu metadanych jest standard Dublin Core, jednak metadane mogą być importowane z innych formatów: XML (m.in. MARCXML, RDF), MARC21, BibTeX.

*dLibra* współpracuje z protokołem OAI-PMH umożliwiającym wyszukiwanie rozproszone w sieciach repozytoriów. Oprócz przeszukiwania metadanych system umożliwia przeszukiwanie zawartości obiektów cyfrowych pozwalających na ekstrakcję tekstu, m.in. HTML, TXT, PDF, DjVu, MS Office, OpenOffice.

Platforma współpracuje z kanałami RSS, systemami pojedynczego logowania (*Single Sign-On*) oraz protokołem LDAP służącym uwierzytelnianiu w oparciu o zewnętrzne bazy LDAP.

Oprogramowanie *dLibra* zostało wdrożone w ok. 150 projektach realizowanych w Polsce, na stronie domowej projektu dostępna jest obszerna dokumentacja w języku polskim, a także fora współpracy użytkowników systemu.

Interfejs użytkownika, oprócz wersji polskiej, oferuje tłumaczenia na języki: angielski, czeski, francuski, niemiecki.

*CDS Invenio* [http://cdsware.cern.ch/]

System rozwijany przez CERN Document Server Software Consortium na otwartej licencji GNU General Public License. Do 2006 roku system nosił nazwę *CDSware*, stworzony został na potrzeby budowania i zarządzania kolekcją dokumentów elektronicznych.

Obsługuje największe repozytorium z zakresu nauk ścisłych i technicznych zarządzane przez Europejską Organizację Badań Jądrowych *CERN Document Server* [http://cdsweb.cern.ch] udostępniający ok. 1 mln dokumentów. *OpenDOAR* odnotowuje oprócz *CDS* dziesięć innych instytucji, przede wszystkim uczelni lub jednostek uczelnianych wykorzystujących to narzędzie. *CDS Invenio* spełnia wymagania standardu Open Archives Initiative (OAI), co umożliwia współpracę z protokołem OAI-PMH.

Standardem opisu bibliograficznego dokumentów jest format MARC21 rozpowszechniony w zintegrowanych systemach bibliotecznych. Administrator repozytorium *Eny PWr* zwraca uwagę m.in. na „*doskonały darmowy support*" oraz fakt, iż system jest elastyczny i łatwo konfigurowalny, umożliwia tworzenie regularnych i wirtualnych kolekcji, personalizację, a także pobieranie i wysyłanie danych w wielu formatach [6]. Obsługiwane formaty eksportu danych to: HTML, BibTeX, Dublin Core, EndNote, NLM, MARC, MARCXML. Interfejs posiada 20 wersji językowych, w tym wersję polską.

## ASPEKTY PRAWNE

W Polsce, w różnym zakresie, prowadzenie działalności związanej z udostępnianiem treści elektronicznych w Internecie może wymagać zgodności z kilkoma ustawami. W ekspertyzie prawnej [16] wykonanej na potrzeby *Kujawsko-Pomorskiej Biblioteki Cyfrowej* (*KPBC*), jej autorka wskazuje, że biblioteka akademicka współtworząca bibliotekę cyfrową powinna realizować swoje zadania w zgodzie z:

- *Ustawą o prawie autorskim i prawach pokrewnych* z dnia 4 lutego 1994 r. (Dz.U. 2000, nr 80, poz. 904 ze zm.)
- *Ustawa o ochronie baz danych* z dnia 27 lipca 2001 r. (Dz.U. 2001, nr 128, poz. 1402)

- *Ustawą o bibliotekach* z dnia 27 czerwca 1997 r. (Dz.U. 1997, nr 85, poz. 539 ze zm.)
- ustawą *Prawo o szkolnictwie wyższym* z dnia 27 lipca 2005 r. (Dz.U. 2005, nr 164, poz. 1365, ze zm.)

W dalszej kolejności istotne mogą się okazać rozporządzenia i wytyczne stosownych resortów, a także szczegółowe zarządzenia i regulaminy obowiązujące w instytucjach tworzących repozytoria i biblioteki cyfrowe, a także regulaminy obowiązujące w instytucjach zatrudniających autorów, których prace są w nich rozpowszechniane.

Przy ustalaniu zasad postępowania z utworami, które zamierzamy wprowadzić do bc lub re, zasadnicze znaczenie ma *Ustawa o prawie autorskim i prawach pokrewnych*.

Ustawodawca określił dla podmiotu prawa autorskiego dwa rodzaje praw: **autorskie prawa osobiste** zapewniające m.in. niezbywalne prawo do autorstwa utworu oraz **autorskie prawa majątkowe**, w myśl których twórcy przysługuje wyłączne prawo rozporządzania utworem oraz otrzymywania wynagrodzenia za korzystanie z niego, to prawo może podlegać zbyciu.

Autorskie prawa majątkowe są obecnie chronione przez 70 lat liczonych od roku następującego po roku śmierci twórcy, a w przypadku utworów, do których z mocy ustawy autorskie prawa majątkowe powstają na rzecz innego podmiotu (np. wydawcy dzieł zbiorowych: encyklopedii, numerów czasopism), „*70-letni termin ochrony jest wyznaczany od daty rozpowszechnienia utworu, a gdy utwór nie został rozpowszechniony – od daty jego ustalenia*" [1].

Znaczna część obiektów udostępnianych w bibliotekach cyfrowych pochodzi ze zbiorów własnych bibliotek, digitalizacji podlegają utwory opublikowane, dla których z mocy ustawy wygasły autorskie prawa majątkowe i utwory przeszły do tzw. domeny publicznej.

Inną grupą obiektów rozpowszechnianych przez bc i re są utwory podlegające ochronie prawnoautorskiej, umieszczone za zgodą autora lub podmiotu, któremu przysługują autorskie prawa majątkowe w wyniku umowy o korzystanie z utworu, często na podstawie licencji niewyłącznej. Specyficznym typem takiego upoważnienia mogą być licencje Creative Commons, które choć niedookreślone na gruncie prawa polskiego [1], bywają stosowane w Polsce przez autorów i wydawców [7] [18]. Autor posiadający prawa majątkowe publikując utwór określa rodzaj licencji na jakiej może być on rozpowszechniany, zastrzegając tym samym pewne prawa. Wybór przez autora takiej formy licencjonowania w praktyce zwalnia instytucję zarządzającą repozytorium z zawierania indywidualnych umów licencyjnych, jeśli zakres działalności repozytorium nie wykracza poza prawa przyznane rodzajem licencji. Rodzaje licencji i zasady ich stosowania prezentuje strona Creative Commons Polska [http://creativecommons.pl/] a praktyczne aspekty i kontekst stosowania omówiono w jednym z referatów [18].

Szczególnym rodzajem utworów, które mogą występować w bibliotekach cyfrowych i repozytoriach są utwory pracownicze, tzn. stworzone w ramach obowiązków wynikających ze stosunku pracy, choć autorskie prawo majątkowe powstaje pierwotnie na rzecz twórcy, to jednak *Ustawa* stanowi, że z chwilą przyjęcia takiego utworu pracodawca nabywa do niego prawa, lecz w granicach określonych zawartą umową o pracę [1].

Korzystniejsze dla twórców reguły przyjęto w *Ustawie* w odniesieniu do utworów pracowniczych, które dodatkowo są utworami naukowymi tworzonymi w instytucji naukowej w ramach obowiązków wynikających ze stosunku pracy, autorskie prawa majątkowe pozostają przy twórcy, a *Ustawa* ogranicza uprawnienia pracodawcy do:
- pierwszeństwa publikacji utworu,
- korzystania z informacji zawartych w utworze,
- udostępniania dzieła osobom trzecim [1].

W ramach uprawnień przysługujących pracodawcy może on zawiesić wykonywanie przysługującego mu prawa np. do pierwszeństwa publikacji utworów lub uzyskiwania przez twórcę zgody pracodawcy na opublikowanie utworu w czasopismach o obiegu międzynarodowym oraz w materiałach konferencyjnych (*Regulamin korzystania z własności intelektualnej Politechniki Wrocławskiej*) [24].

Prawa do całości wydawnictw zbiorowych, w tym artykułów opublikowanych w czasopismach, przynależą wydawcy. Wedle przywołanej powyżej ekspertyzy prawnej dla *KPBC*: „*wydawca lub producent uzyskuje »ex lege« (na mocy prawa), w sposób pierwotny prawa do całości dzieła... Przyjmuje się, że dzieło zbiorowe istnieje tylko wówczas, gdy autorzy przynajmniej w sposób dorozumiany aprobują uczestniczenie w pracach związanych z tworzeniem dzieła*" [16]. Według innej interpretacji „*by pozbawić twórcę majątkowych uprawnień autorskich ze względu na eksploatację utworu w dziele zbiorowym, konieczne jest zawarcie z nim stosownej umowy*"[1].

Autorzy zamierzający umieścić w bc lub re utwory, które opublikowali wcześniej w czasopismach naukowych powinni sprawdzić treść umów zawieranych z wydawcami, które w wielu przypadkach wymagają definitywnego transferu autorskich praw majątkowych lub w inny sposób mogą ograniczać autorowi korzystanie z utworu w wersji przed recenzją naukową, jak i wersji ostatecznej, zatwierdzonej do druku.

Zasady obowiązujące przy publikacji w czasopismach ok. 600 wydawców naukowych oraz możliwościach dysponowania utworami pozostawionych twórcom można sprawdzić w bazie danych powstałej w projekcie SHERPA/RoMEO [sherpa: http://www.sherpa.ac.uk/romeo/].

Zdeponowanie przez autora utworu w bc lub repozytorium z reguły wymaga złożenia przez niego wraz z utworem podpisanej umowy licencyjnej umożliwiającej jego wykorzystanie z określonym umową zakresie. Jednak niektóre z projektów przywołanych przez inną analizę sporządzoną na potrzeby *KPBC* [17], prezentującą rozwiązania brytyjskie, niemieckie i holenderskie w zakresie stosowania prawa autorskiego w procesie digitalizacji i tworzenia bibliotek cyfrowych, wskazują, że w kilku projektach holenderskich, autorzy mogą w trakcie rejestracji na stronie internetowej repozytorium akceptować warunki licencji niewyłącznej poprzez kliknięcie przycisku „Register", bez konieczności przekazywania licencji papierowej.

## BIBLIOTEKI CYFROWE I REPOZYTORIA NA RZECZ NAUKI

Według wszystkich deklaracji (Budapest, Bethesda, Berlin) konstytuujących nowy model komunikacji w nauce, jedną z dróg rozwoju Open Access jest tworzenie repozytoriów. W coraz większym stopniu decyzja o wyborze sposobu publikacji wyników badań naukowych jest niezależna od decyzji autora, ale podyktowana jest wymaganiami stawianymi przez grantodawców np. National Institutes of Health [9] lub pracodawców np. wydziały uniwersytetów Harvarda i Stanforda [5]. Instytucje nakładają obowiązek na autorów prowadzących badania z funduszy rządowych, aby publikowali wyniki badań w czasopismach Open Access lub innych czasopismach umożliwiających autorom – równoległe lub opóźnione o umowny okres – umieszczanie w repozytoriach materiałów w wersji preprint lub postprint.

W polskich warunkach funkcje repozytoriów instytucjonalnych, w pewnym stopniu, przejęły biblioteki cyfrowe, które udostępniają oprócz skanowanych zbiorów zgromadzonych w bibliotekach znajdujących się w domenie publicznej, także współczesne publikacje własnych pracowników i doktorantów, m.in. kolekcje prac doktorskich, książki i numery czasopism wydawane w oficynach uczelnianych oraz materiały konferencji organizowanych na uczelniach.

Opracowania o charakterze naukowym z założenia występują w repozytoriach. Biblioteki cyfrowe mają bardziej uniwersalny charakter, oprócz aktualnych prac naukowych udostępniają materiały dydaktyczne, dokumenty archiwalne, materiały o znaczeniu historycznym i kulturalnym, dokumenty życia społecznego, prasę, w dużej mierze regionalną.

Niezależnie od charakteru prezentowanych w bc zbiorów, wszystkie one mogą stać się przedmiotem badań naukowych różnych dyscyplin.

Szczególne znaczenie dla prowadzenia bieżących prac badawczych mogą mieć czasopisma i książki przekazywane przede wszystkim przez oficyny uczelniane, a także doktoraty. W grupie współczesnych czasopism naukowych reprezentowanych w *DBC* odnaleźć można m.in. ostatnie roczniki: „Argumenta Oeconomica", „Advances in Clinical and Experimental Medicine", „Dental and Medical Problems", „Materials Science", „Polymers in Medicine", „Śląskiego Przeglądu Statystycznego" i „Wiadomości Chemicznych". Podkolekcję doktoratów odnotowuje zarówno *BCUWr*, jak i *DBC*, w której pięć uczelni prezentuje ten typ dokumentów. Książki pochodzące z własnych oficyn udostępnia sześć uczelni współpracujących z *DBC*, jednak ograniczenia natury prawnej i niewielkie zainteresowanie ze strony autorów sprawiają, że tylko część produkcji wydawniczej oficyn trafia do bc.

## WNIOSKI

W tworzenie bc i re na Dolnym Śląsku włączyły się przede wszystkim biblioteki szkół wyższych oraz biblioteki publiczne. Dostrzegalny zaczyna być udział innych jednostek uczelnianych tworzących nowe projekty lub włączających swoje zbiory do istniejących (wydziały, instytuty, oficyny wydawnicze, archiwa).

Rozproszone projekty realizowane przez instytucje regionu, dzięki wykorzystaniu standardowych protokołów pobierania metadanych, tworzą kompleksową ofertę czytelniczą.

Projekty związane z gromadzeniem i udostępnianiem kolekcji cyfrowych wymagają długoterminowego zabezpieczenia finansowego. Praktykowanym powszechnie rozwiązaniem jest poszerzanie działalności instytucji o ten aspekt i przejmowanie odpowiedzialności za utrzymanie i rozwój projektów przez jednostki organizacyjne instytucji. Początkowe koszty uruchomienia bc lub re mogą nie być znaczące, zwłaszcza jeśli wykorzystują one aplikacje typu *open source*, wymaga to jednak wykwalifikowanej kadry. Uwzględnić trzeba również rosnące potrzeby sprzętowe i profesjonalne aplikacje do obróbki dokumentów elektronicznych.

W odniesieniu do obu prezentowanych platform zgłosić należy pewne zastrzeżenia związane z uzyskaniem informacji statystycznych. Interfejs użytkownika w *dLibrze* pozwala na śledzenie szczegółowych informacji m.in. o kształtowaniu się całej kolekcji w czasie, jej wykorzystaniu, liczbie odwiedzin i wyszukiwań, typach zasobów (formie wydawniczej i piśmienniczej), formacie danych, języku, prawach własności; nie umożliwia jednak śledzenia wykorzystania i kształtowania się kolekcji poszczególnych instytucji, a zebranie tych danych poprzez opcję „Najczęściej czytane", przy tak dynamicznie rozwijających się kolekcjach, jest zadaniem karkołomnym. *CDS Invenio* nie posiada specjalnego modułu statystycznego dostępnego dla użytkownika. Ogólne informacje takie, jak liczba obiektów, liczba i zawartość kolekcji znajdują się na stronie startowej. Być może więcej informacji pozwala uzyskać panel administratora, jednak ani czytelnik, ani potencjalny autor nie jest w stanie uświadomić sobie jak pożyteczne może być repozytorium, i że warto zdeponować w nim swoje prace.

Natomiast umieszczenie przez autorów prac w repozytoriach wpływa na wcześniejsze udostępnienie prac, wyższe wykorzystanie oraz większą liczbę ich cytowań [10].

Łatwiejszy dostęp do kolekcji elektronicznych w Internecie zapewnia stosowanie przez bc i re uniwersalnych standardów, z którymi współpracują również popularne wyszukiwarki, wpływa to na wykorzystanie zbiorów i rozpoznawalność instytucji, a także buduje pozytywny wizerunek instytucji i przyczynia się do promowania autorów.

Lepszej „widzialności" w sieci sprzyja również rejestrowanie bc i re w globalnych serwisach i katalogach agregujących informacje o nich i ich metadanych tworząc w ten sposób uniwersalną wirtualną metabibliotekę.

## SUMMARY
### DEVELOPMENT OF DIGITAL LIBRARIES AND ELECTRONIC REPOSITORIES IN LOWER SILESIA IN YEARS 2004-2008

In following elaboration were presented digital libraries and electronic repositories operating in Lower Silesia region in years 2004-2008. General description of character and size of their collections was presented, as well as standards and methods of digital collections management and juridical aspects of this management. Potential of usage of digital collections in regional scientific researches was described.